# Physical Design of Superconducting Magnet for ADS Injection I

PENG Quan-ling(彭全岭), WANG Bing(王冰), CHEN Yuan(陈沅) YANG Xiang-chen(杨向臣)
Institute of High Energy Physics, Chinese Academy of Sciences, Beijing 100039, China

**Abstract**

Superconducting solenoid magnet prototype for ADS Injection I had been designed and fabricated, and also had been tested in liquid Helium state inside a vertical Dewar in Haerbin institute of Technology in Nov. 2012. The design current is 210A, when the test current reached 400A no quench occurred, the solenoid magnet was forced to quench by the embedded heaters. The integral field strength, leakage field at the nearby upstream and downstream superconducting spoke cavities all meet the design requirements. At the same time, it also checked the system reliability of the vertical test Dewar and the quenched detection system. The superconducting magnet prototype has accumulated valuable experiences for the coming batch magnets production and cryogenic test.

## 1 Introduction

The Accelerator Driven Subcritical System (ADS), by using proton beam hitting on the target to produce neutron for spent fuel stable increment or accelerating evolution, aims at looking key solutions for nuclear fuel stable provision and for the safe nuclear waste treatment [1]. Figure 1 is the schematic overview of ADS linac accelerator. The injection I and the main linac will be designed and fabricated by IHEP. The spoke 325MH part in injection I have two cryomodules, each one consists of 6 superconducting spoke cavities, 5 superconducting magnets, and 5 beam position monitors connected along the beam line in series. Figure 2 shows the layout of these Accelerator components inside the cryomodule. The proton beams will be accelerated from 3Mev at the outlet MEBT1 to 5MeV by these 6 SC spoke cavities. Each superconducting magnet, which indeed is a magnet package aimed for beam focusing and orbit correction, contains a solenoid magnet, a horizontal dipole corrector (HDC) and a vertical dipole corrector (VDC). The integral field strength for the solenoid is 0.4T.m, the maximum integral field for HDC and VDC is 1600Gs.cm.

* Supported by China ADS Project (XDA03020000)

1) E-mail: pengql@ihep.ac.cn

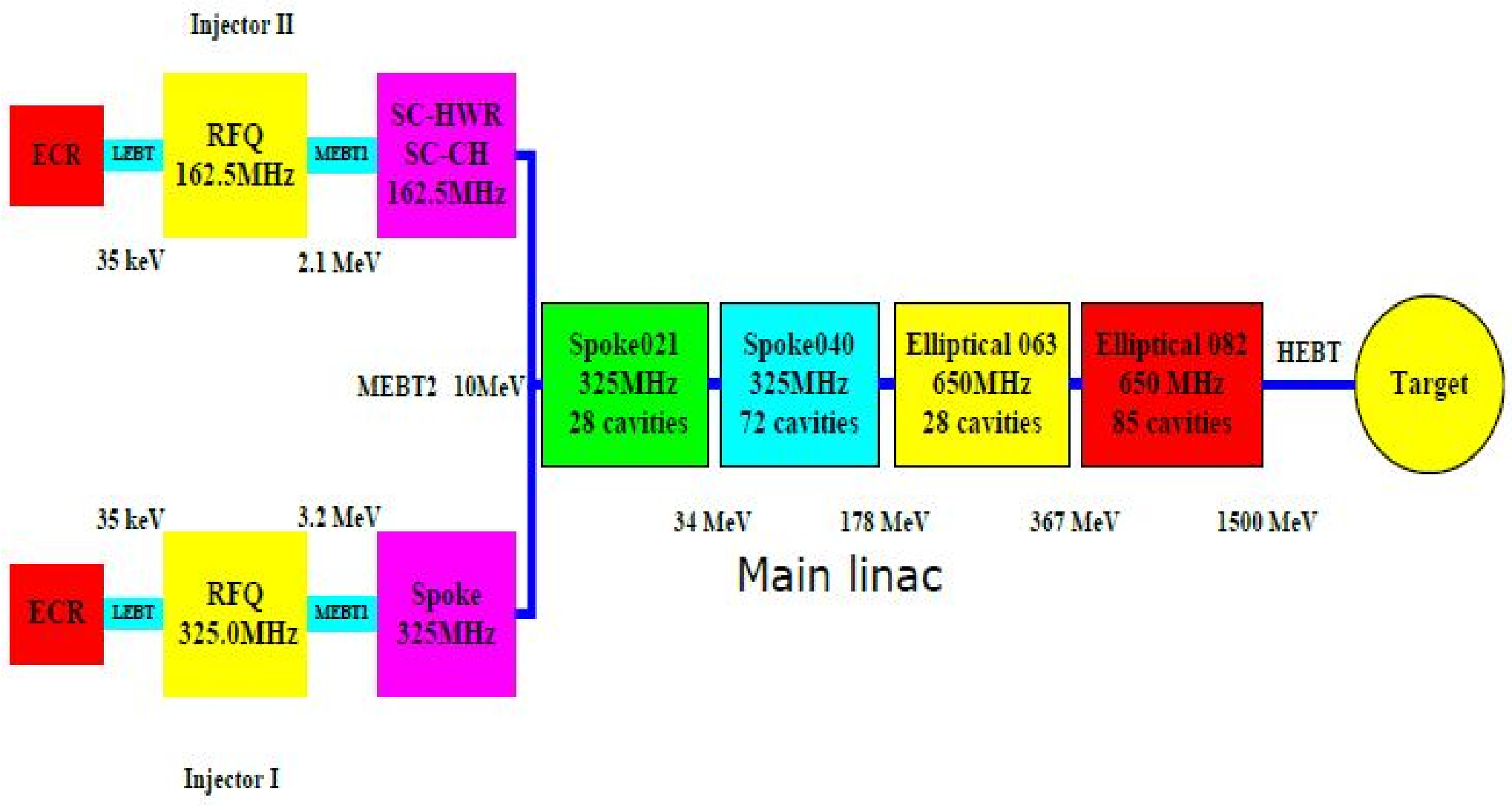

Fig 1 Schematic overview of ADS linac accelerator

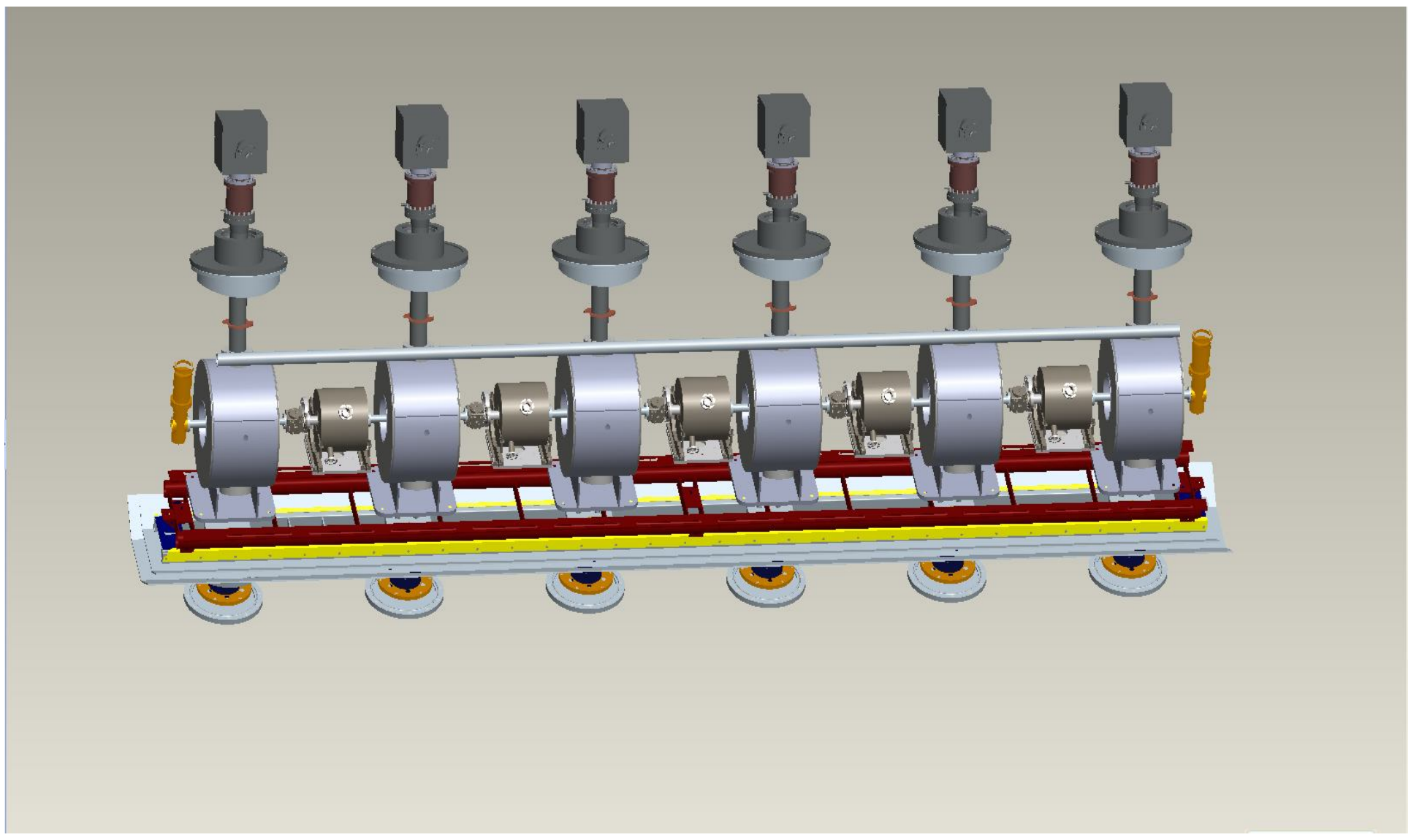

Fig 2 Accelerator components inside the first cryomodule (Along the beam line, the big one represents the superconducting spoke cavity, the middle one is the superconducting magnet, the small one is the beam position monitor)

**2 Physical design of the superconducting magnet prototype**

There is a SC spoke cavity each at upstream and downstream of the superconducting magnet. Leakage field, mainly coming from the solenoid, will add with the earth magnetic field and then affect the operation quality of the spoke cavity or even drive it to quench. The design aim for the leakage field reduction is to realize a less than 1Gs field at the distance of 270mm away from the solenoid center. The SC spoke cavity, with the extra magnetic shield made of high permeability perm-alloy covered, the field inside the cavity can be further reduced to less than 0.1Gs.

At the design stage, three methods are hired to reduce the leakage field from the solenoid, they are: 1) using three solenoids with main solenoid in the middle and a bucking solenoid at each end of the main solenoid to compensated the tail field; 2) added iron return yoke surround these three solenoids to form the field return path; 3) with a high permeability perm-alloy tube covering the return yoke to reduce the small leakage field more efficiently. 2D and 3D OPERA [2] software are used to optimize the field calculation, Figure 3 is the final 2D field model, Figure 4 is the 3D calculation model. At the spoke cavity place with 270mm away from the solenoid center, the leakage axial field are $4.5\times10^{-6}$T with perm-alloy shield and $3\times10^{-5}$ T without per-malloy shield. The axial repulsive forces two bucking solenoids bore are 4.7kN, which must be eliminate by preload forces during the magnet assembly.

The SC cable used for the solenoid coils are Chinese domestic productions, the SC:Cu ratio is 4:1， the critical current is 700A at 6T. The design operation current is 210A, which has a high safety margin.

Saddle shaped coil will be selected for HDC and VDC, they save spaces and then reduce the magnet energy. Figure 5 shows the 3D field calculation for the correctors. The coil first wound in a flat pattern, then wrapped around the support tube into a saddle shape. Superconducting wire with 0.35mm in diameter and 0.67:1 for SC:Cu is used for winding the correction coils. Their design currents are 12A, the critical current is 100A at 3T.

Calculation methods in reference [3] are used, the setting threshold for the quench detection time is 50ms. Taking the solenoid magnet coil as an example, the maximum hot spot temperature inside the coil is 40K as shown in Table 1. Including the quench action time, total quench protection time should be less than 60ms.

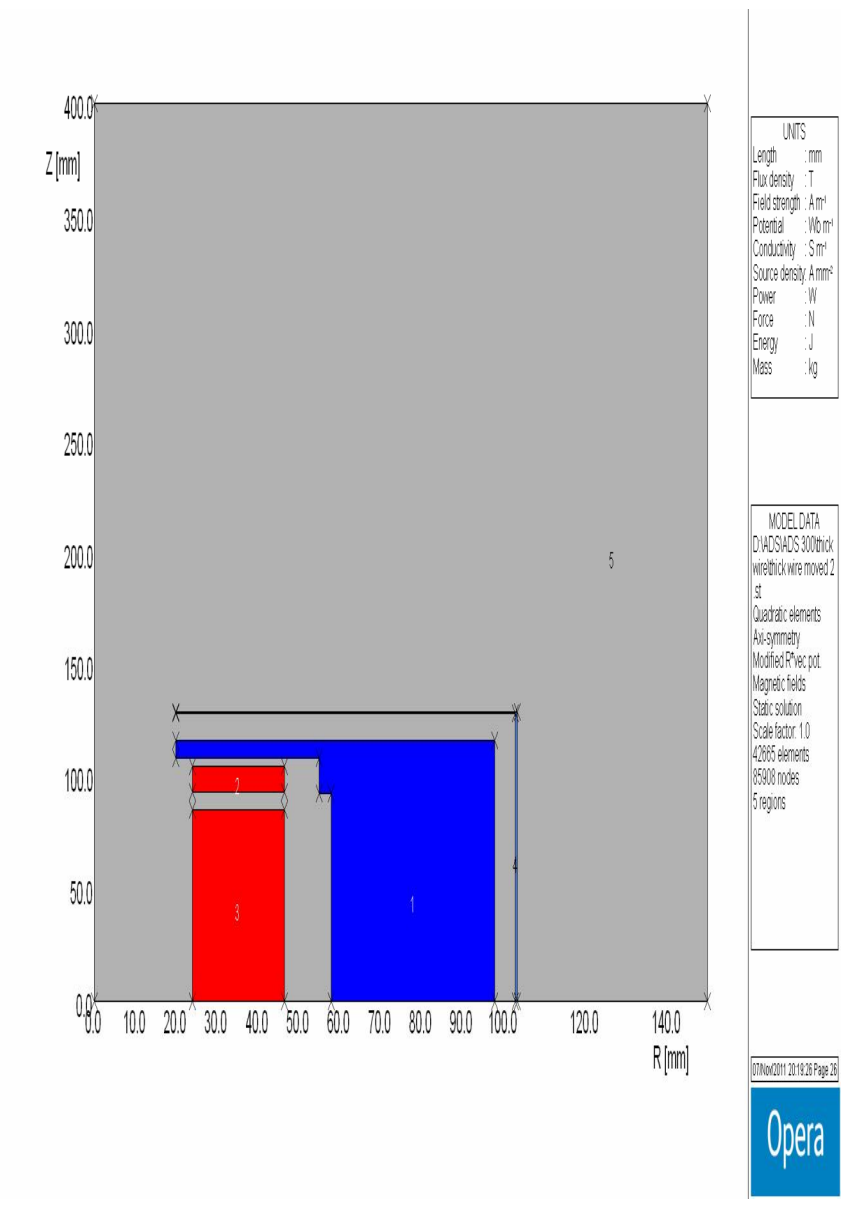

Fig 3 2D calculation model for the solenoid field

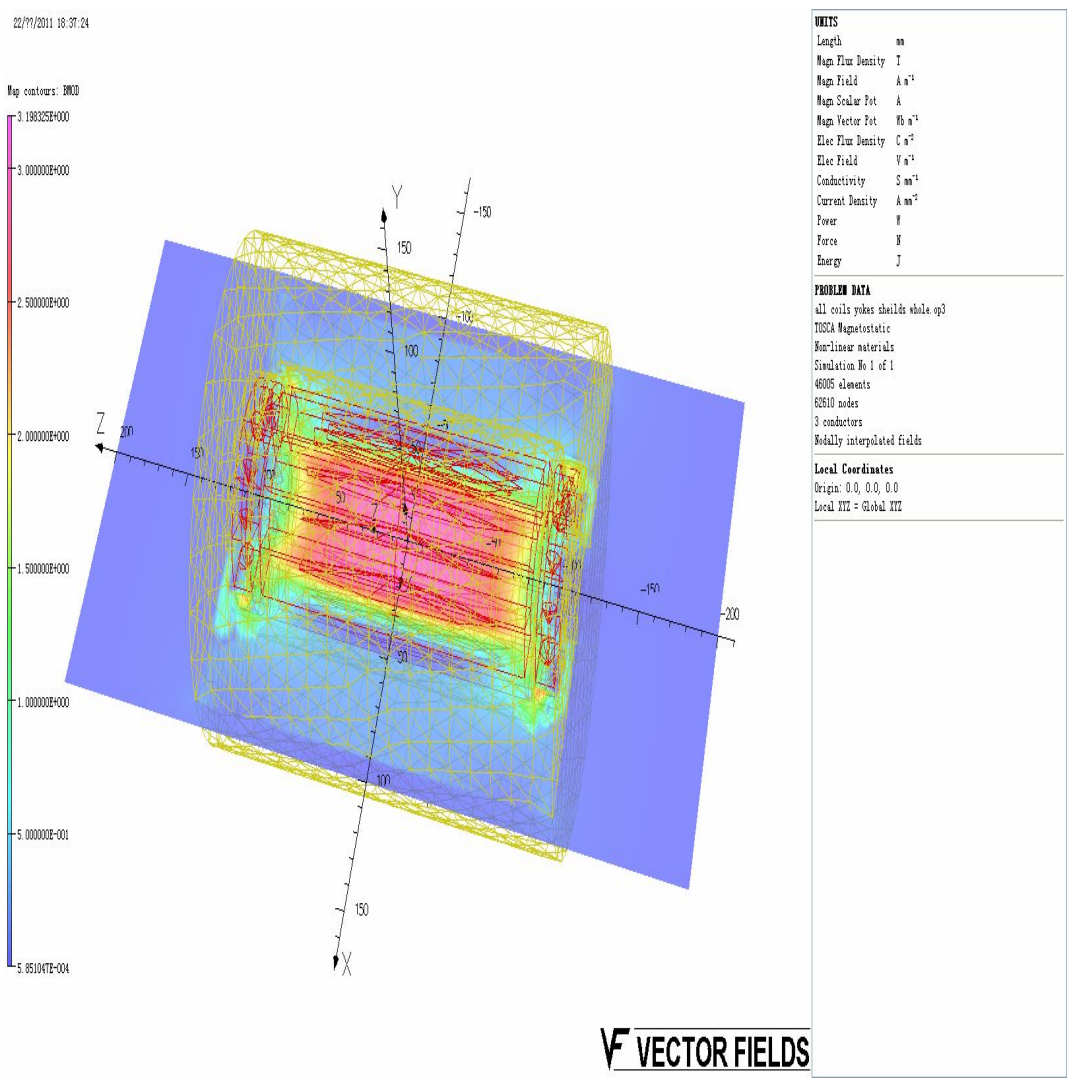

Fig 4 3D Magnetic field calculation for the solenoid field

Table 1 Design Parameters for ADS superconducting magnet prototype

| | Main Solenoid | Bucking solenoid | HDC and VDC correctors |
|---|---|---|---|
| Dimension for SC cables | 1.2×1.8 $mm^2$ | 1.2×1.8 $mm^2$ | ϕ 0.35mm |
| Cu:SC ratio | 4:1 | 4:1 | 0.67:1 |
| Maximum field on the coil | 3.3 | -- | 3.2 |
| Total turns | 2208 | 182×2 | 60 |
| layers/turns per layers | 24/92 | 26/7 | 1/30 |
| Operation current (A) | 210 | 210 | 12 |
| Coil length (mm) | 170.2 | 13 | -- |
| Current density J0 ($A/mm^2$) | 110 | 110 | 170 |
| Store Energy $E_0$（kJ) | 2.6 | 0.02 | 0.002 |
| Inductance (mH) | 120 | 1.2 | 25 |
| $E_0J_0^2$ ($A.J/m^4$) | $6.0\times10^{20}$ | --- | --- |
| $F(T_{max})$ ($A.s/m^4$) | $1.19\times10^{16}$ | -- | --- |
| Hot spot temperature（K） | 40 | --- | -- |

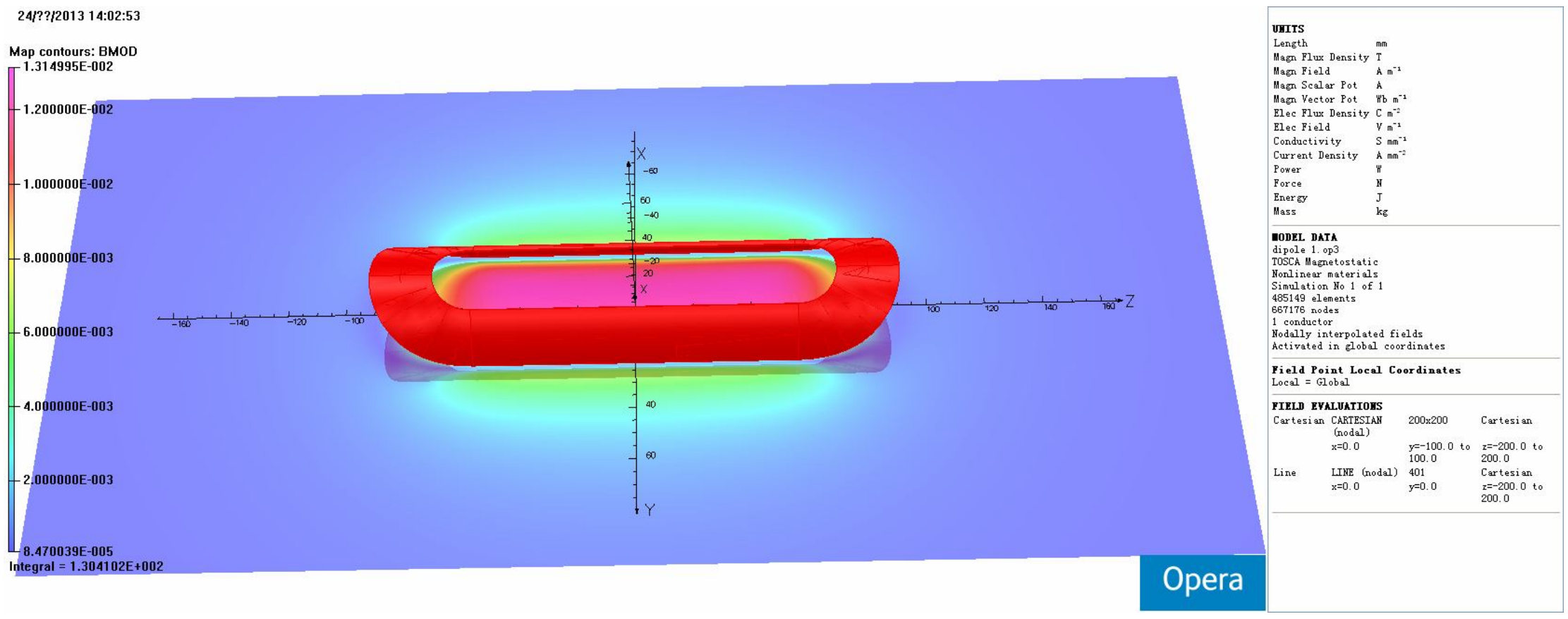

Fig 5 3D magnetic field calculation for the corrector

## 3. Magnet fabrication and vertical test

Including the end connection flanges, total length for the magnet is only 300mm。In order to connect the upstream BPM flange, an asymmetric structure is selected for the magnet cryostat, the liquid helium vessel is 15.8mm shifted to the spoke cavity side. The magnet cryostat is put inside the whole vacuum space of the cyomoudle, no liquid nitrogen shield is needed; beam vacuum chamber is also used as the coil support tube and as the inner liquid helium vessel. Figure 6 shows the mechanical cross section view of the magnet cryostat。For easy shaping and fixing, the perm-alloy shield is immersed in the liquid helium space.

The bare magnet, where the outer liquid vessel is not welded on, will be test in 4.2K liquid helium state for quench test and field measurement. Figure 7 shows the superconducting magnet hanged at the bottom of the test stand, here the whole top flange set are taken out from the vertical test Dewar in order for the magnet installation and signal wires connection.

Active quench protection scheme is selected for ADS superconducting magnet. Quench test was the first process after enough liquid helium was accumulated inside the vertical Dewar feed by a cryogenic system. The excited current for the solenoid magnet successfully reached 400A after several current steps without quench. No quench occurred after kept the solenoid at 400A for 10 minutes, then the solenoid was forced to quench by electrical heaters embedded in the inner layers of the main solenoid coil, the quench detection system was triggered in time, the informed power supply was cut off almost simultaneously at the same time, then an 1.0 ohm dump resister was put into the quench protection circuit to keep the magnet in safety. All the two corrector magnets reached 30A without quench, which are much higher than the design current.

Figure 8 shows the magnet ongoing vertical test process, the field measurement platform was installed on the top of the vertical Dewar. The moving module, which sits on the top of the vertical test Dewar, pulls a 3D hall probe moving in an isolated warm bore tube in every 1mm step vertically, the maximum moving distance is 460mm. For test results, the solenoid integral field is 0.3810T.m at 210A, which is smaller than the design value, since the measured rod where the hall probe put in is not enough to cover the tail of the solenoid field. For leakage field, including the earth field, the axial field is only 0.8Gs at 200mm away from the magnet center, which is much lower than the requirement. The field at the spoke cavity place, the field will be dropped much lower. More detailed measurement process will be done in the horizontal test.

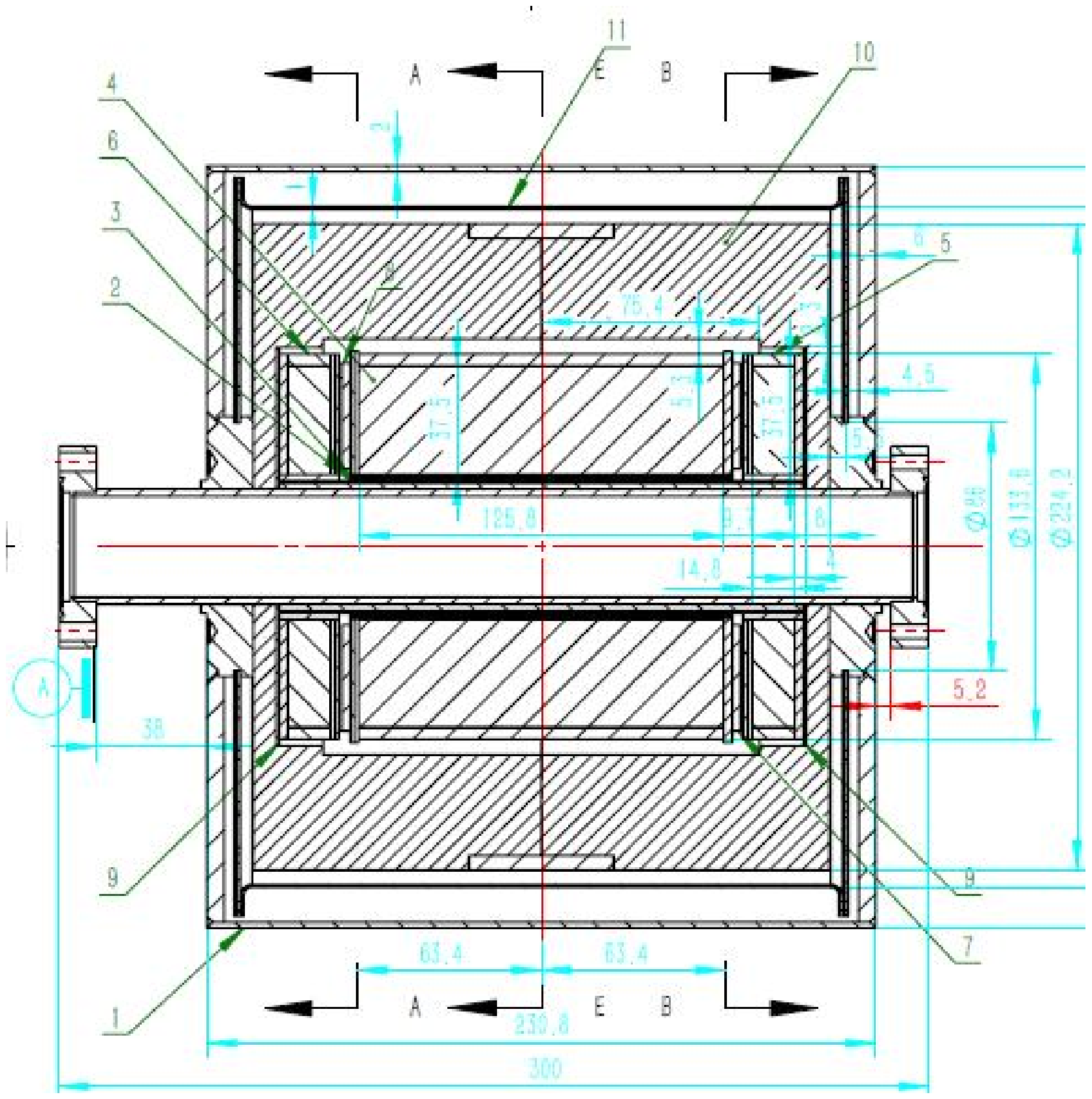

Fig 6 Cross section overview for the Superconducting magnet cryostat

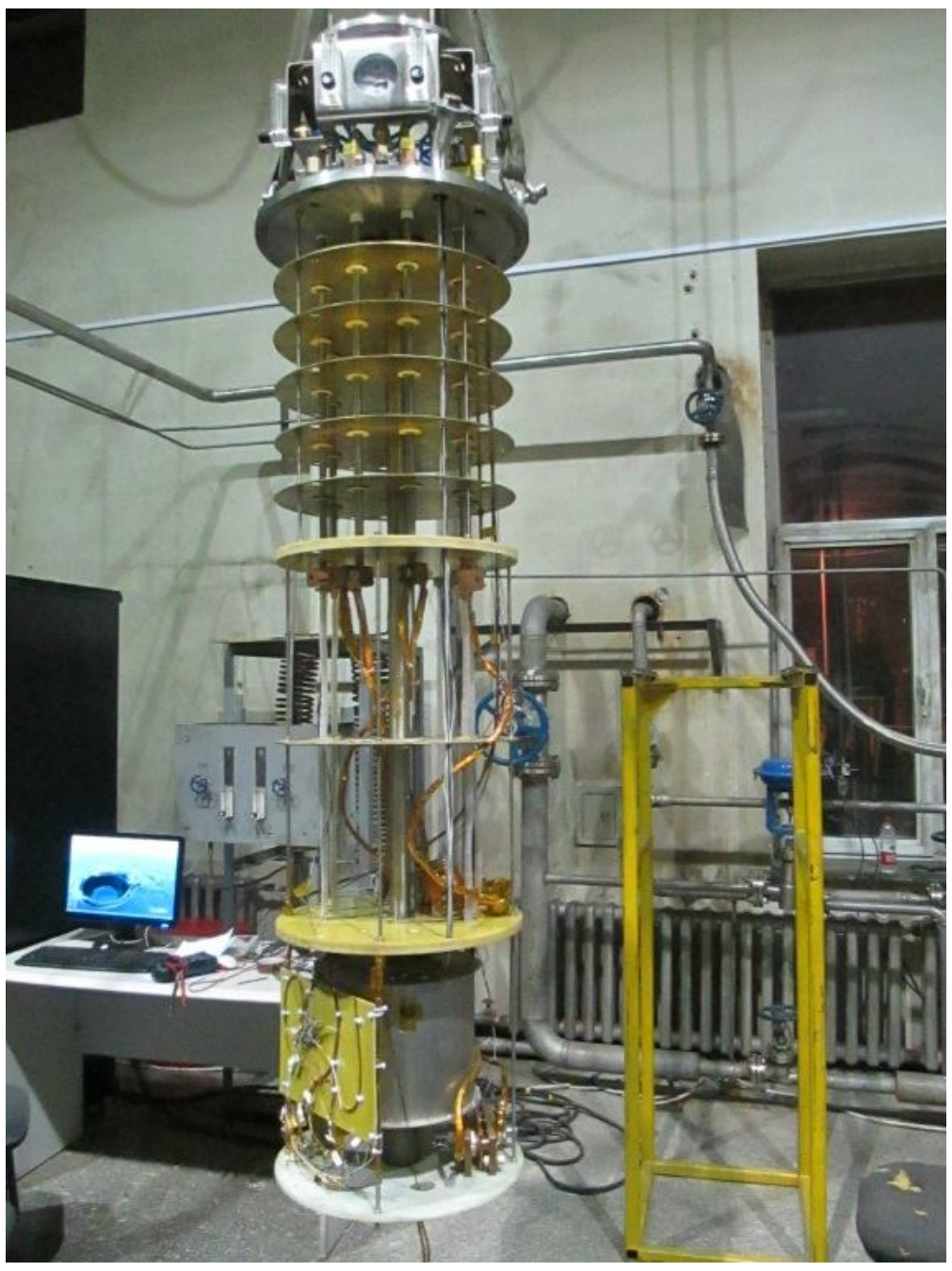

Fig 7 Superconducting magnet installed at the bottom of the vertical test stand, the superconducting joints are welded and fixed on the G10 board

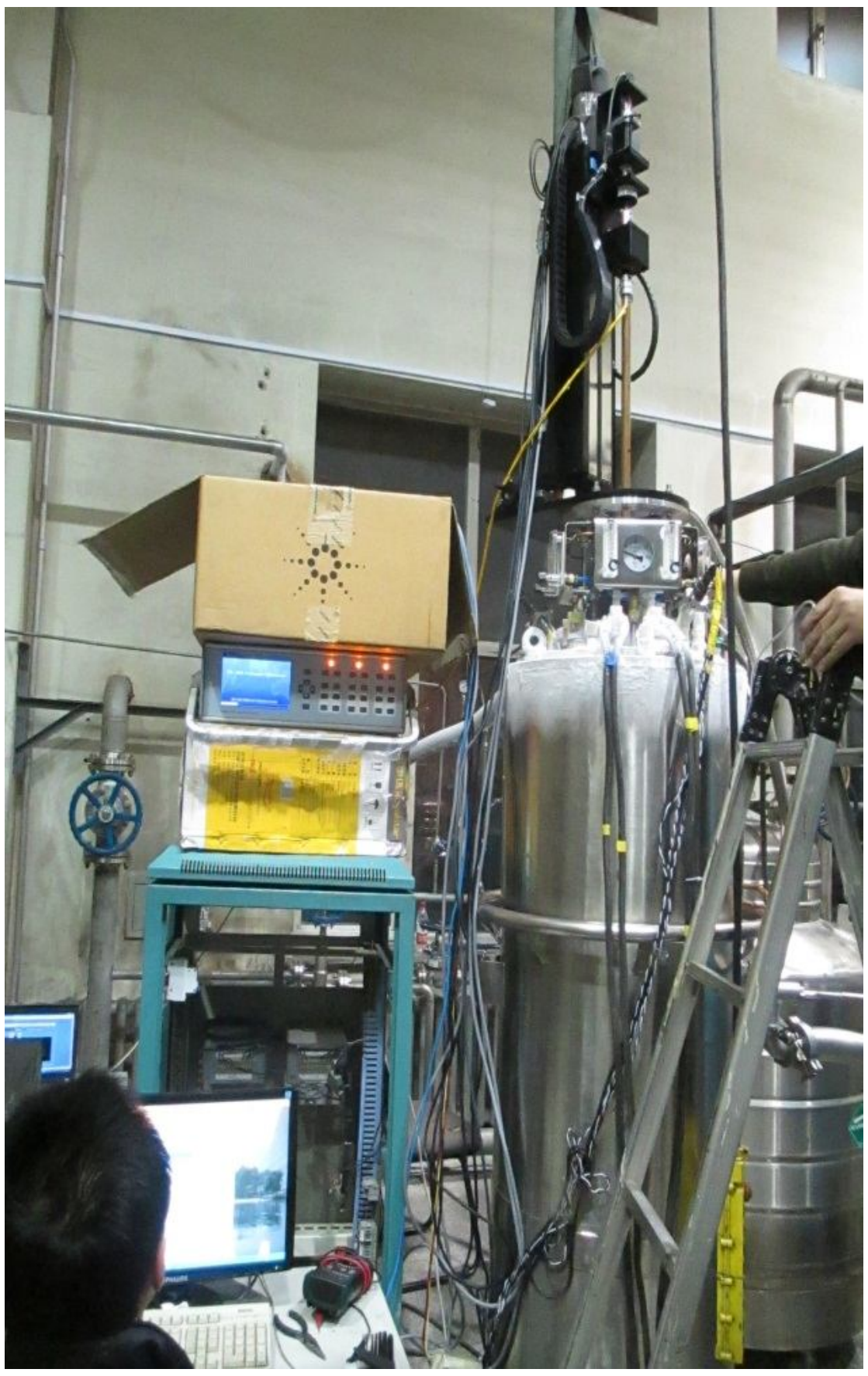

Fig 8 Ongoing vertical test for the superconducting magnet

**4. Design improvement for the batch magnet production**

Total 12 SC magnets will be fabricated and test for injection I. From the prototype magnet design, coil winding, installation and vertical test, several problems should be solved before the batch magnet production. The perm-alloy magnetic shield tube will be removed off since it is not easily machined and assembled, small place between the iron yoke and the per-alloy shield is difficult for SC cables and signal wires pulling through. For a better leakage field reduction effect, the iron yoke will be extended to the previous per-alloy place for more iron to confine the leakage field. Figure 9 shows the new design scheme, form the 2D calculation results it is better than that of the prototype magnet. 3D magnetic calculation was done for consideration of the wholes on the return yoke for cables and cryogenic channels; the used calculation model is shown in Figure 10. The SC cable for the solenoid will be used a small dimension with $0.55\times0.85mm^2$ for reason of to reduce the heat load to the cryogenic system. The main solenoid coil was shortened and the peak field was increased to 4.5T, but is still in a high safety margin.

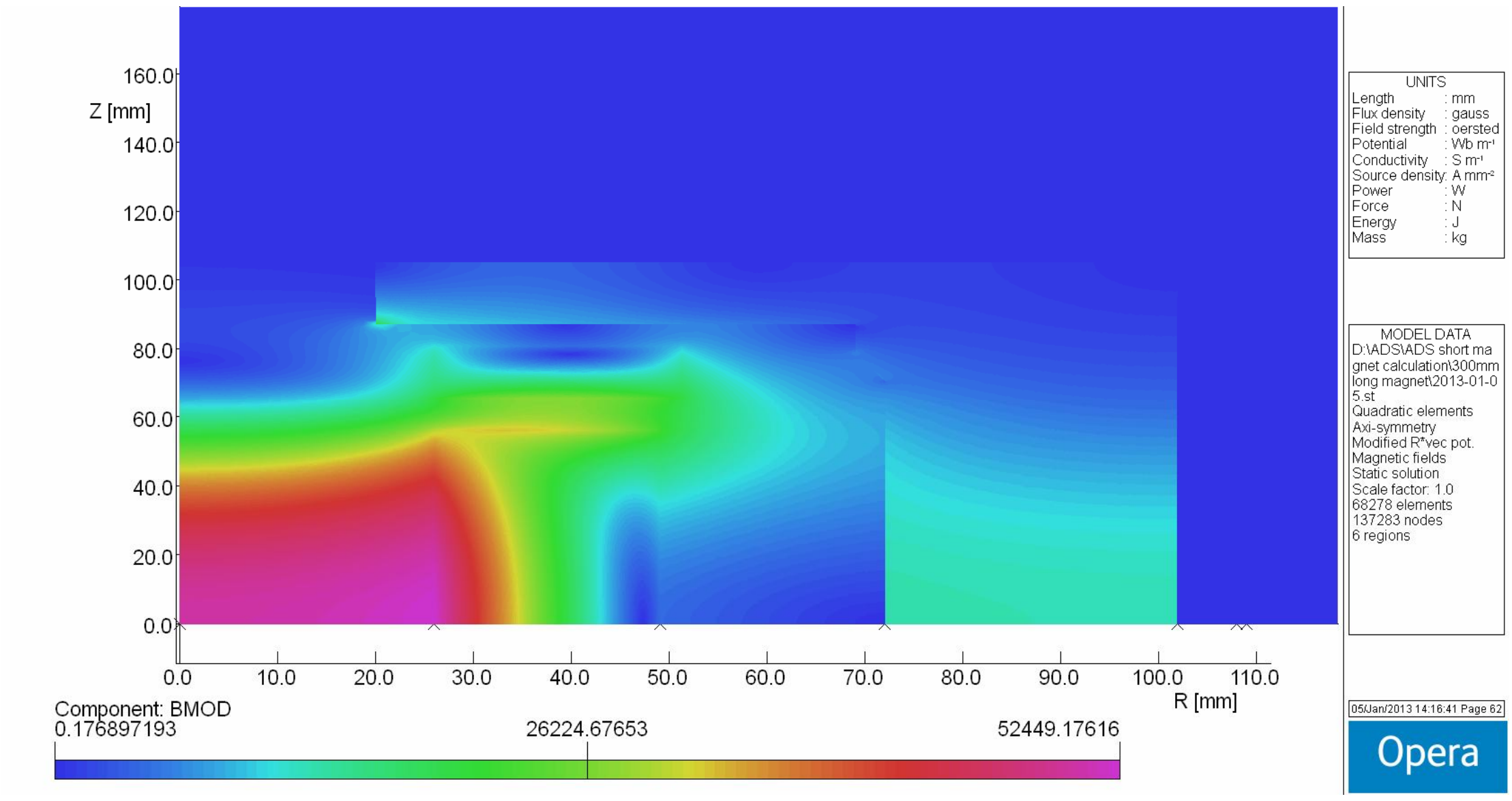

Fig 9 2D field profile for the batch magnet, the calculate current is 20% higher than the design value.

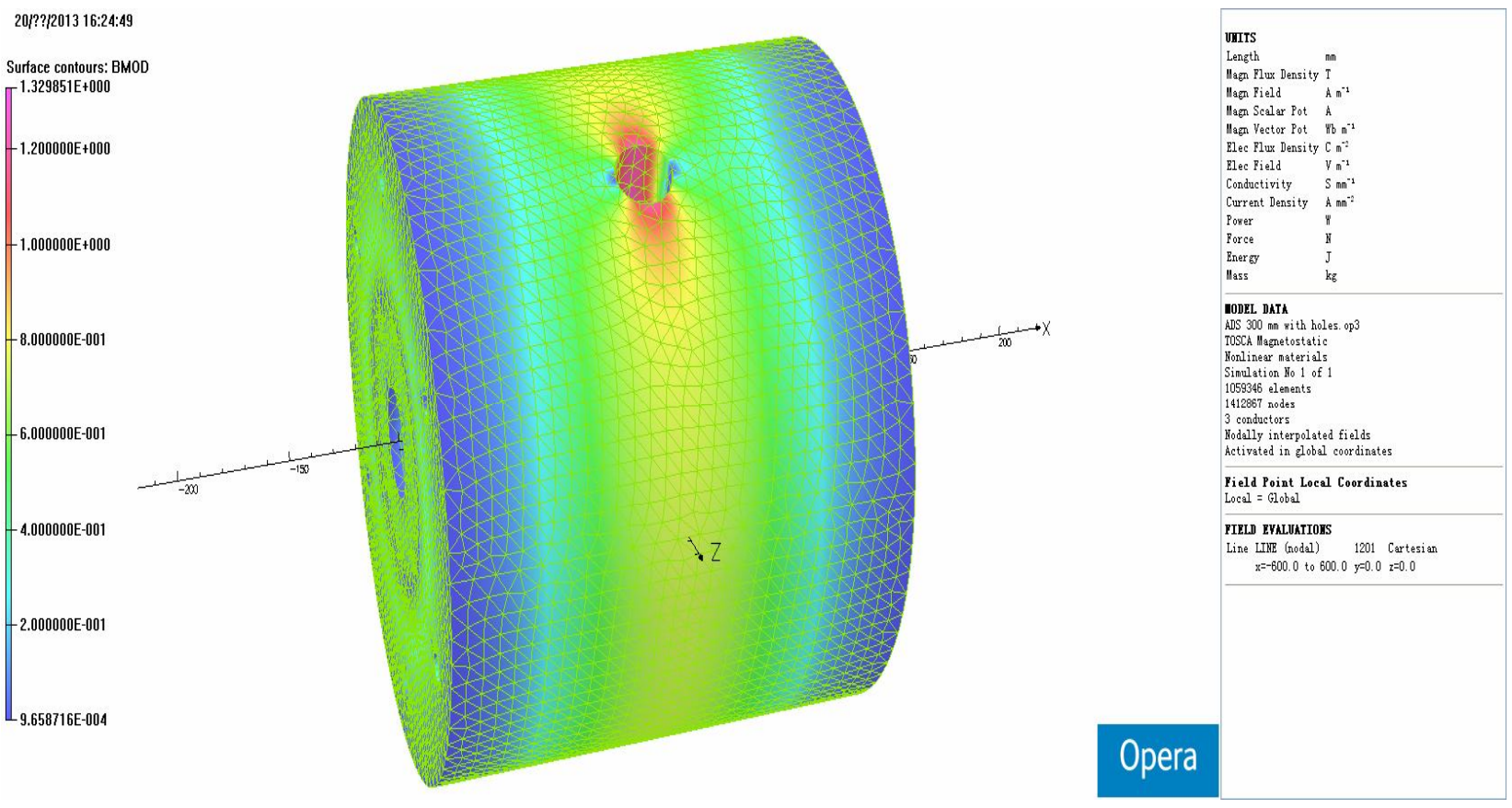

Fig 10 3D field calculation model for the batch magnet

## 5. Conclusion

A superconducting magnet prototype for ADS injection I linac has been fabricated and test. System reliability of the vertical test Dewar and the quenched detection system has been checked at the same time. It has accumulated valuable experiences and directions for the batch magnets production and cryogenic test. Some modifications will be taken for future batch magnet production, which includes, reduce the design current, remove the per-alloy shield and prolong the iron yoke.